\documentclass{article}

\usepackage{arxiv}

\usepackage{amsmath}
\usepackage{amsfonts}
\usepackage{graphics}
\usepackage{amsmath,amssymb}
\usepackage[utf8]{inputenc} % allow utf-8 input
\usepackage[T1]{fontenc}    % use 8-bit T1 fonts
\usepackage{hyperref}       % hyperlinks
\usepackage{url}            % simple URL typesetting
\usepackage{booktabs}       % professional-quality tables
\usepackage{amsfonts}       % blackboard math symbols
\usepackage{nicefrac}       % compact symbols for 1/2, etc.
\usepackage{microtype}      % microtypography
\usepackage{lipsum}
\usepackage{mathtools}
\usepackage{here}
\usepackage{graphicx}
\usepackage[affil-it]{authblk}
\usepackage{comment}

\title{\textbf{Emergence of Human-Like Attention in Self-Supervised Vision Transformers: an eye-tracking study}}

\author[1]{Takuto Yamamoto}
\author[2]{Hirosato Akahoshi}
\author[1, 2, 3]{Shigeru Kitazawa\thanks{Corresponding author. Email: \texttt{kitazawa@fbs.osaka-u.ac.jp}}}
\affil[1]{Department of Brain Physiology, Graduate School of Medicine, The University of Osaka, 1-3 Yamadaoka, Suita, Osaka 565-0871, Japan.}
\affil[2]{Dynamic Brain Network Laboratory, Graduate School of Frontier Biosciences, The University of Osaka, 1-3 Yamadaoka, Suita, Osaka 565-0871, Japan.}
\affil[3]{Center for Information and Neural Networks (CiNet), National Institute of Information and Communications Technology, 1-4 Yamadaoka, Suita, Osaka 565-0871, Japan.}
\date{}
\begin{document}

\makeatletter
\newenvironment{tablehere}
  {\def\@captype{table}}
  {}

\newenvironment{figurehere}
  {\def\@captype{figure}}
  {}
\makeatother
\columnseprule=0.5mm
\columnsep=2em
\maketitle

\newcommand{\argmax}{\mathop{\rm arg~max}\limits}
\newcommand{\argmin}{\mathop{\rm arg~min}\limits}

\begin{abstract}
Many models of visual attention have been proposed so far. Traditional bottom-up models, like saliency models, fail to replicate human gaze patterns, and deep gaze prediction models lack biological plausibility due to their reliance on supervised learning. Vision Transformers (ViTs), with their self-attention mechanisms, offer a new approach but often produce dispersed attention patterns if trained with supervised learning. This study explores whether self-supervised DINO (\textit{self-DIstillation with NO labels}) training enables ViTs to develop attention mechanisms resembling human visual attention. Using video stimuli to capture human gaze dynamics, we found that DINO-trained ViTs closely mimic human attention patterns, while those trained with supervised learning deviate significantly. An analysis of self-attention heads revealed three distinct clusters: one focusing on foreground objects, one on entire objects, and one on the background. DINO-trained ViTs offer insight into how human overt attention and figure-ground separation develop in visual perception. 
\end{abstract}

\keywords{attention\and eye-tracking\and vision transformer\and self-supervised learning\and DINO\and figure ground separation}

% \begin{multicols}{2} \end{multicols}
\section{Introduction}
Biological visual systems use attention to efficiently extract information from the environment. Traditional models of bottom-up attention, such as saliency models, highlight salient points based on low-level visual features like edges, color, and luminance, using hand-crafted filters that mimic early visual processing \cite{Harel2006-wp,Itti1998-py,Itti2001-qh}. However, these models, such as the Graph-based Visual Saliency (GBVS) model \cite{Harel2006-wp}, have limitations. For instance, GBVS failed to replicate human gaze patterns when viewing video clips \cite{Suda2015-he}. \par
Recently, convolutional neural networks (CNNs) have been used to predict human gaze using supervised learning on human gaze data \cite{Cornia2018-ul,Kroner2020-ur,Kummerer2017-dy}. While these deep gaze prediction models perform well, they lack biological plausibility, as human attention is learned without explicit instruction. In biological systems, attention serves visual perception, meaning that attention models must be integrated with broader models of visual perception \cite{Salehinajafabadi2024-jr, Lei2021-mu}. \par
In this context, the vision transformer (ViT), introduced by Dosovitskiy, et al. \cite{Dosovitskiy2021-tp}, is notable for its use of attention mechanisms in image classification. ViT divides an image into patches, which are treated as tokens analogous to those used in the original transformer model developed for natural language processing \cite{Vaswani2017-ho}. In addition to the patch tokens, ViT uses a class token to aggregate information across patches via self-attention, which connects to a classification head. However, the supervised training of the original ViT reduces its biological plausibility, and it often produces noisy, unfocused attention. \par
Caron, et al. \cite{Caron2021-qj} addressed this by training ViTs with the self-supervised DINO method (\textit{self-DIstillation with NO labels}), which produced more focused attention confined within object boundaries, closely resembling human-like attention. However, this observation lacks quantitative validation. \par
\begin{figure}[h]
  \centering
  \includegraphics[clip]{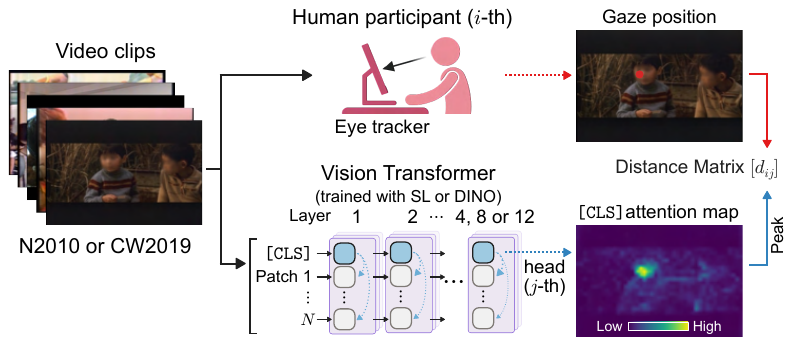}
  \caption{\textbf{Comparison of gaze coordinates between human participants and attention heads of vision transformers (ViTs).} Video clips from N2010 (Nakano et al., 2010) \cite{Nakano2010-sg} and CW2019 (Costela \& Woods, 2019) \cite{Costela2019-ba} were presented to ViTs. The gaze positions of each self-attention head in the class token (\texttt{[CLS]}), identified as peak positions within the self-attention map directed at patch tokens, were compared with human gaze positions from the respective datasets. There were six ViT models, varying by the number of layers ($L = 4,8,$ or $12$) and training methods: supervised learning (SL) or self-supervised learning using the DINO method.}
  \label{fig:fig1}
\end{figure}

In this study, we quantitatively compared human overt attention, measured through eye-tracking, with the attention generated by ViTs trained under two conditions: DINO and conventional supervised learning (Fig. \ref{fig:fig1}). To enhance ecological validity, we used video stimuli, as human gaze behavior remains remarkably consistent across individuals during dynamic viewing \cite{Nakano2010-sg}. We show that DINO-trained ViTs closely match the overt attention of neurotypical adults, while ViTs trained using conventional methods diverge significantly. Additionally, an analysis of individual self-attention heads revealed three distinct clusters: one focusing on the ``center'' of foreground objects (e.g., a face), another on entire objects (e.g., bodies), and a third on background areas. We discuss these findings in relation to psychological and physiological attention theories. 

%\begin{comment}
%\end{comment}

%%%%%%%%%%%%%
%% Results %%
%%%%%%%%%%%%%
\section{Results}
\subsection{DINO ViTs exhibit attention like TD adults}
\begin{figure}[p]
  \centering
  \includegraphics[clip, width=\linewidth]{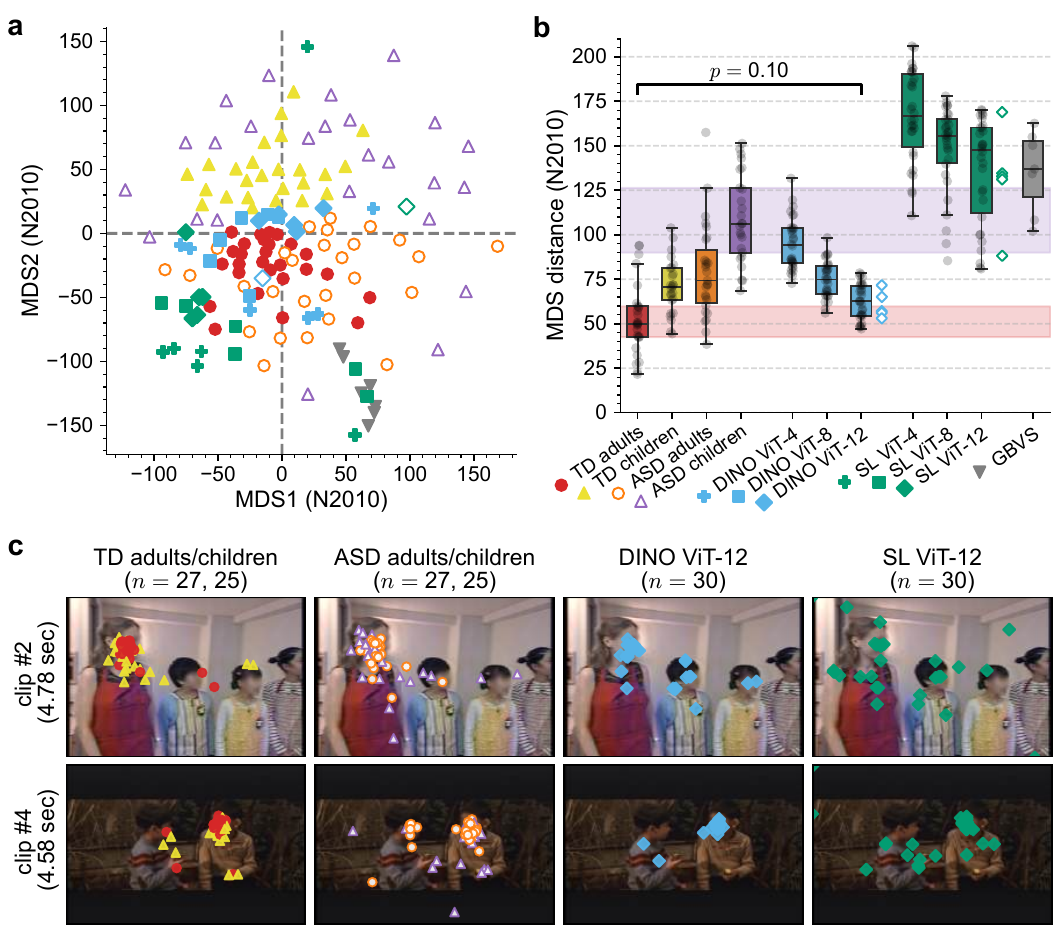}
  \caption{\textbf{Comparison of temporo-spatial gaze patterns in humans and ViTs while viewing N2010 video clips using multidimensional scaling (MDS).} \textbf{a}, Gaze patterns of human participants, ViTs, and Graph-Based Visual Saliency (GBVS) models, are displayed on a two-dimensional MDS plane. Symbols represent data from four participant groups from N2010 (red dots: adults with typical development (TD), yellow triangles: TD children, orange circles: adults with autism spectrum disorder (ASD), purple triangles: ASD children), six ViT models (cyan symbols: DINO ViTs, green symbols: SL ViTs), and the GBVS models (gray inverted triangles). Open diamonds show performance of the official models, DeiT-S trained with SL (green diamond), and ViT-S/16 trained with DINO (cyan diamond). Note that TD adults (red dots) and DINO ViTs (cyan symbols) are distributed near the center, while SL ViTs (green symbols) and GBVS (gray) are in the periphery. Only the best head of each ViT is plotted. \textbf{b}, Comparison of MDS distance from the origin in the 32-dimensional MDS space across 10 groups (human, ViT, and GBVS). Data are shown for the top-5 heads. Symbols for each group, as used in (\textbf{a}), are shown below. Horizontal shadings indicate the interquartile ranges for TD adults (red) and ASD children (purple). Notably, DINO ViT-12 is the only group with MDS distance comparable to TD adults. All comparisons between TD adults and other groups were significantly different ($p<0.001$, Wilcoxon rank-sum tests, two-sided, Bonferroni corrected) except for the comparison with DINO ViT-12 (indicated with a bracket and $p$-value). \textbf{c}, Examples of gaze points for human participants and ViTs while viewing clip No.2 (``Okaasan to Issho'', NHK) and clip No.4 (``Always: Sunset on Third Street'', Toho Co., Ltd). For the ViTs, data from the top-5 heads are shown. }
  \label{fig:fig2}
\end{figure}

We projected the temporo-spatial ``gaze'' patterns of 36 ViTs onto a two-dimensional MDS plane determined by landmark gaze profiles from human participants (Fig. \ref{fig:fig2}a). These landmark gaze data, collected from 104 human participants in a previous study (N2010) \cite{Nakano2010-sg}, were based on a 77-second video composed of 12 short clips featuring varying numbers of human characters. The same video was presented frame-by-frame to the ViTs---18 trained with DINO (cyan symbols) and 18 trained with supervised learning (SL; green symbols)---and the best head of each ViT was plotted on the MDS plane.\par
The origin of the MDS plane represents the median of all human gaze patterns. Adults with typical development (TD adults, red circles, $n = 27$) clustered near the origin, indicating high consistency in their gaze profiles, which serve as the standard for human gaze patterns. Notably, the DINO ViTs, particularly the 8- and 12-layer models (cyan squares and diamonds, six for each), were distributed near the origin, closely resembling the TD adults. In contrast, SL ViTs (green symbols) were distributed farther from the center, even more distant than adults with ASD (orange circles, $n = 27$), indicating a larger divergence from both TD adults and human participants in general. Additionally, the gaze profiles predicted by graph-based visual saliency (GBVS) models were located in the most distant region of the plane (gray inverted triangles), further underscoring the limitations of traditional models.\par
To gain a more comprehensive view of these temporo-spatial gaze profiles, we increased the MDS dimension from 2 to 32, positioning the top five heads of each ViT in the 32-dimensional MDS space. We then quantitatively compared the MDS distance, defined as the distance from the origin of this 32-dimensional space, across four human participant groups, six DINO and SL ViT groups, and the GBVS model (Fig. \ref{fig:fig2}b). The MDS distances for TD adults were concentrated within the smallest range, serving as the benchmark against which other distributions were compared (horizontal shading in red, Fig. \ref{fig:fig2}b). Remarkably, the 12-layer DINO ViT (DINO ViT-12) was the only group whose MDS distances were statistically comparable to those of TD adults ($p=0.10$, Wilcoxon rank-sum test, after Bonferroni correction). All other groups exhibited significantly greater MDS distances ($p<0.001$). \par
Two additional points merit attention. First, SL ViTs produced substantially larger MDS distances compared to DINO ViTs, despite both having identical transformer architectures ($p<0.0001$, in all 9 comparisons). This result suggests that DINO significantly outperforms conventional supervised learning in replicating human-like attention. Second, MDS distances decreased as the number of layers increased from 4 to 8 and 12.\par
These results are further illustrated in Fig. \ref{fig:fig2}c, using two typical frames from the video. In clip No. 2, TD participants focused their gaze on the face of the woman on the left, while in clip No. 4, their gaze was directed to the face of the boy on the right. In contrast, participants with ASD displayed a broader distribution of attention. Similarly, the 12-layer DINO ViTs (DINO ViT-12) directed their attention to the same faces as TD participants, whereas SL ViTs-12 exhibited more dispersed attention, covering faces, bodies, and other areas of the scene.\par

\subsection{Gaze pattern similarity generalizes to drama but less to animation or nature films}
\begin{figure}[p]
  \centering
  \includegraphics[clip, width=0.99\linewidth]{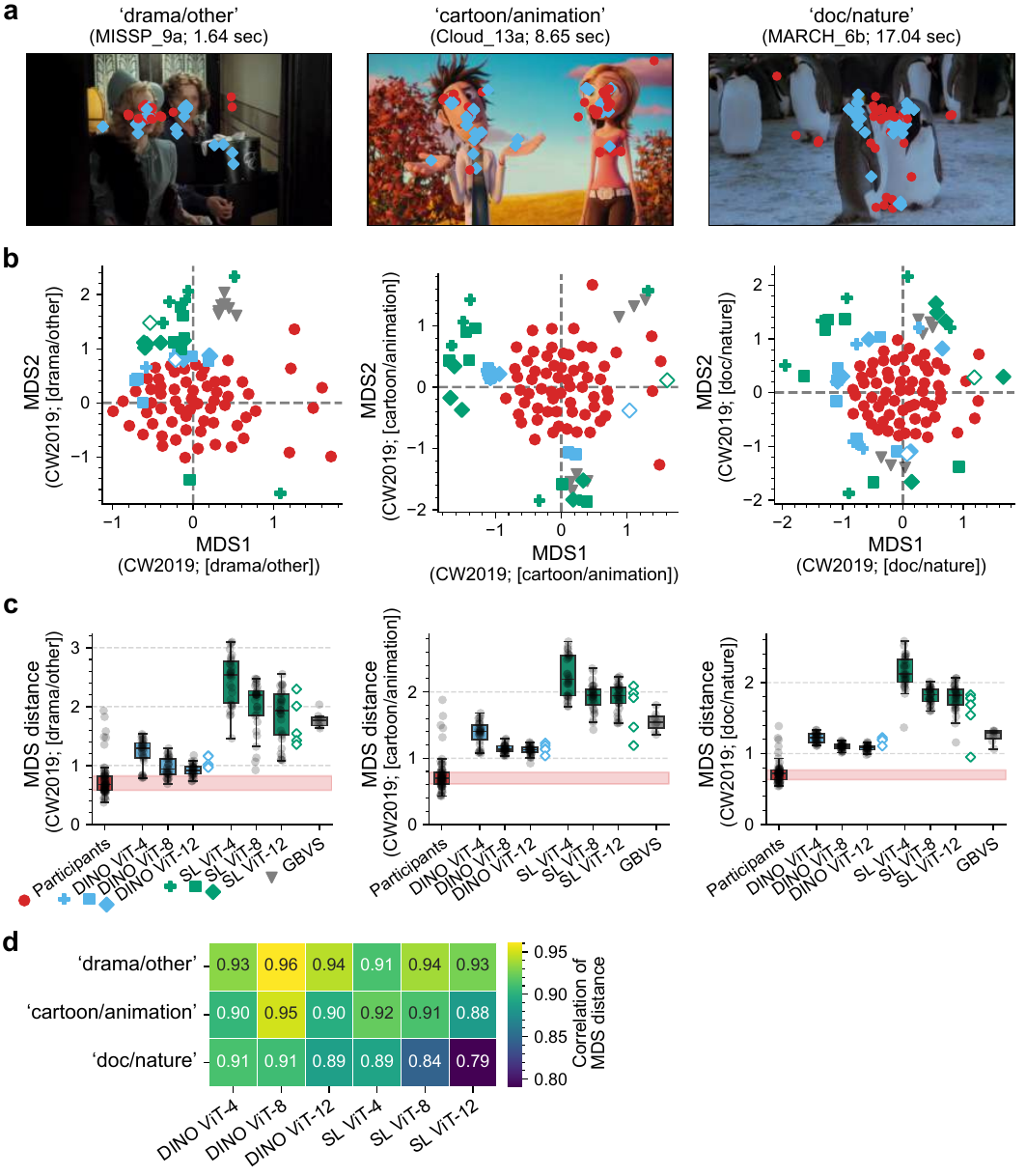}
  \caption{\textbf{MDS analysis of gaze patterns on the CW2019 dataset.} \textbf{a}, Gaze positions of human participants (red dots) and DINO ViTs-12 (cyan diamonds, top-5 heads of six ViTs) plotted on a typical frame from the three video genres: drama/other (left, ``Miss Pettigrew Lives for a Day'', Focus Features LLC and Momentum Pictures), cartoon/animation (middle, ``Cloudy with a Chance of Meatballs'', Columbia Pictures Industries, Inc. and Sony Pictures Entertainment, Inc.), and documentary/nature (right, ``March of the Penguins'', National Geographic Films). \textbf{b},\textbf{c}, Gaze patterns plotted on the MDS planes (\textbf{b}) and group comparison of MDS distance (\textbf{c}), with each panel showing data from a single video genre. Conventions follow those in Fig. \ref{fig:fig2}a and b. Note the similarity between the results for the drama/other genre here and those for N2010 shown in Fig. \ref{fig:fig2}b, both of which featured human characters. \textbf{d}, Pearson correlation coefficients between MDS distances calculated from the N2010 and CW2019 datasets, using all self-attention heads in each model for the calculation.}
  \label{fig:fig3}
\end{figure}

To test generalizability of the above-mentioned MDS analysis, we applied the same approach to a different data set (CW2019 \cite{Costela2019-ba}, 63 healthy adult participants). The dataset contained 200 videos across three genres: ``drama/other'' (featuring human actors), ``cartoon/animation'' (featuring human-like characters) and ``documentary/nature'' films (featuring animals or natural scenes), as illustrated in Fig. \ref{fig:fig3}a.\par
For the drama/other genre, which featured human actors as in N2010, the results were largely consistent with the findings with N2010. The adult participants formed a cluster in the center of the MDS plane, with DINO ViTs positioned close by, while SL ViTs and GBVSs were located farther from the center (Fig. \ref{fig:fig3}b and c, left panels). However, discrepancies emerged between the attention patterns of human participants and DINO ViTs for the cartoon/animation genre. These discrepancies became even more pronounced in the documentary/nature genre without human characters. Conversely, the performance of GBVS predictions approached the human cluster when the documentary/nature films were used. This suggests that human attention may rely more on visual saliency cues when no human characters are present in the scene. \par
The extent of generalization was further quantified by calculating correlation between the MDS distances obtained from N2010 and those from each of the three genres in CW2019 (Fig. \ref{fig:fig3}d). As expected, the correlation was highest for the drama/other genre (0.91--0.96), followed by cartoon/animation (0.88--0.95), and documentary/nature films (0.79--0.91). To conclude, findings with N2010 dataset generalized to those with CW2019 dataset as long as the video clips featured human characters.\par

\subsection{Gaze pattern similarity generally peaks before the final layers}
\begin{figure}[t]
  \centering
  \includegraphics[clip, width=\linewidth]{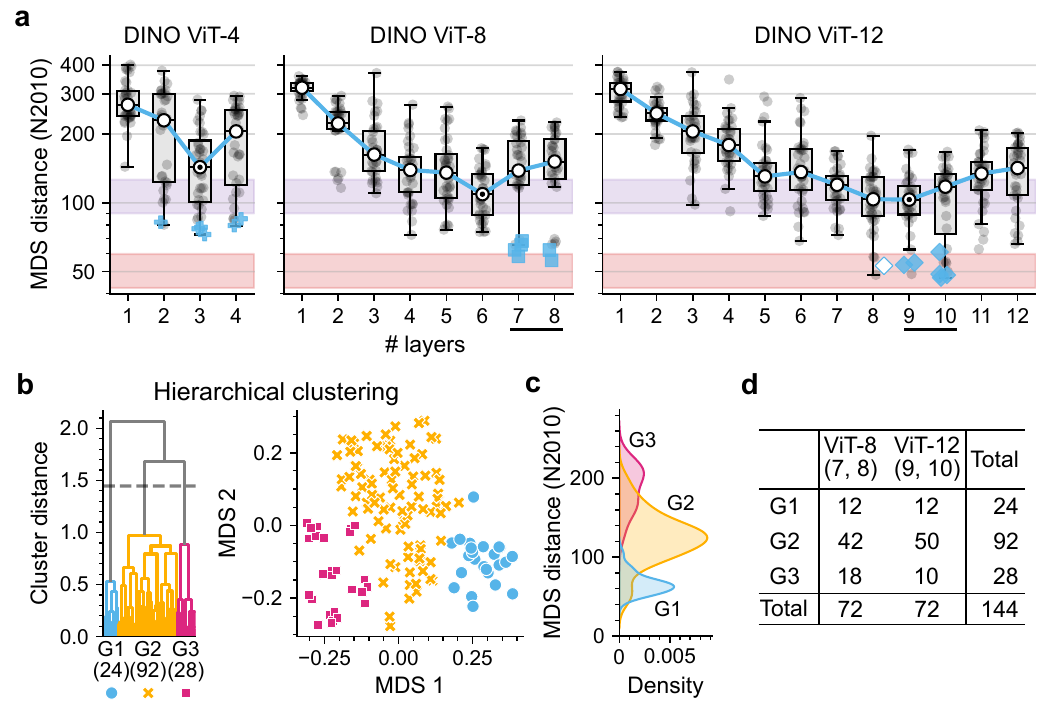}
  \caption{\textbf{Layer-wise MDS distance distribution (N2010) and hierarchical clustering of self-attention heads in DINO ViTs.} \textbf{a}, MDS distance of all self-attention heads at each layer in the DINO ViTs. The white circle represents the median MDS distance per layer, and $\odot$ markers indicate the layer with the lowest median value. The best head in each model is marked by cyan (DINO) or green (SL) symbols, corresponding to the heads shown in Fig. \ref{fig:fig2}a. Bands are consistent with those in Fig. \ref{fig:fig2}b. \textbf{b}, Hierarchical clustering of self-attention heads from the 7th and 8th layers of DINO ViT-8 and the 9th and 10th layers of DINO ViT-12, based on attention maps from N2010 frames. (left) Dendrogram; (right) MDS embedding of the self-attention heads using cosine distance between attention maps. Clustering resulted in three groups: G1, G2, and G3. \textbf{c}, MDS distance distribution (N2010) for heads in G1, G2, and G3. \textbf{d}, Table showing the number of the self-attention heads included in G1, G2, and G3.}
  \label{fig:fig4}
\end{figure}

We then examined how MDS distance changes across different layers of ViTs (Fig. \ref{fig:fig4}a). Each ViT has six attention heads for the class token in each layer, and we analyzed six individual ViTs, giving us a total of 36 attention heads per layer. In DINO ViT-12, the median MDS distance (open circles) decreased from layer 1 to layers 8 and 9, then increased, forming a U-shaped curve. Additionally, the attention heads formed two clusters in layers 9 and 10, with most of the best heads (cyan diamonds) found in the smaller cluster. \par
These observations were generally applied to DINO ViT-4 and DINO ViT-8 as well (Fig. \ref{fig:fig4}a). Similar patterns were found for SL ViTs, although the MDS distances in SL ViTs were significantly larger compared to their DINO ViTs counterparts (Supplementary Fig. \ref{fig:fig_sup1}). These observations remained largely unchanged when using the CW2019 dataset (Supplementary Fig. \ref{fig:fig_sup2}).

\subsection{Three distinct groups of attention heads emerge in DINO ViTs}
To further investigate whether distinct clusters exist in the optimal layers of ViT-8 (layers 7 and 8) and ViT-12 (layers 9 and 10), we analyzed the cosine similarity of the attention maps across the 144 attention heads in these four layers (4 layers $\times$ 6 heads $\times$ 6 models). A hierarchical cluster analysis based on cosine similarity revealed that the distribution was best represented by three clusters (Fig. \ref{fig:fig4}b), as determined by voting across 24 indices using NbClust \cite{Charrad2014-kw}. The G1 heads showed the smallest MDS distance, while the G2 heads formed an intermediate peak, and the G3 heads exhibited the largest peak (Fig. \ref{fig:fig4}c). G1, G2, and G3 consisted of 24, 92, and 28 heads, respectively, out of the 144 heads (Fig. \ref{fig:fig4}d).

\subsection{G1, G2, and G3 heads focus on the face, body, and background}
\begin{figure}[p]
  \centering
  \includegraphics[clip, width=\linewidth]{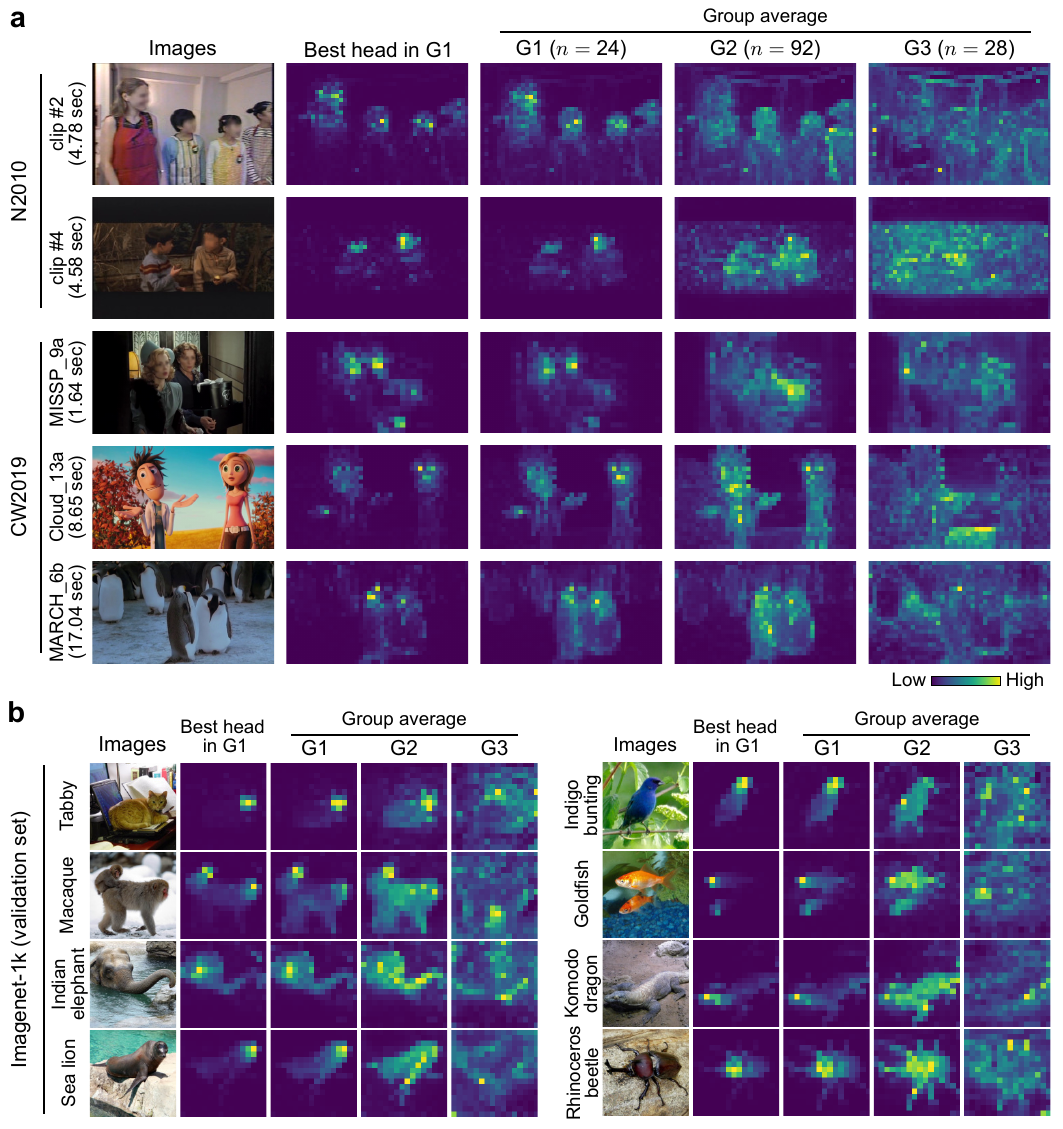}
  \caption{\textbf{Focus areas of the three self-attention head groups—face (G1), body (G2), and background (G3).} \textbf{a}, Attention maps of the best head in G1 (DINO ViT-8, model 2, head 1) and the group-averaged attention maps for G1, G2 and G3 heads across five frames, two from N2010 dataset, and three from CW2019 dataset. These frames are the same as those shown in Fig. \ref{fig:fig2}c and Fig. \ref{fig:fig3}a. \textbf{b}, Attention maps of the best head in G1 (same head as in (\textbf{a})) and the group-averaged attention maps for G1, G2 and G3 across eight animal images from the ImageNet-1k validation set.}
  \label{fig:fig5}
\end{figure}

We compared the attention patterns of the three groups of heads (G1, G2, and G3) on typical frames taken from five video clips (Fig. \ref{fig:fig5}a). As expected from the smallest MDS distance, the G1 heads focused on human faces, or the faces of penguins, with a peak on the face of the primary character in each scene. In contrast, the G2 heads distributed attention over the bodies of the attended characters, forming sharp contours that segregated them from the background. By comparison, the G3 heads focused on the rest, namely the background. Notably, the G3 heads often highlighted vertical and oblique lines that delineated the borders of walls and ceilings (e.g., clip No. 2, MISSP\_9a; G3). \par
Figure \ref{fig:fig5}b compares the attention patterns on images of non-human animals sampled from ImageNet-1k dataset. G1 heads attended to the faces, while the G2 heads attended to the entire body, and the G3 heads focused on the background. This pattern held across various species, including mammals (tabby, macaque, Indian elephant, and sea lion), birds (indigo bunting), fish (goldfish), reptiles (Komodo dragon), and insects (rhinoceros beetle).  

\subsection{G1 heads focus more on faces than TD adults and children}
We further examined the duration of attention (viewing time) that each G1 head allocated to the eyes, mouth, or face (Fig. \ref{fig:fig6}a), similar to the analyses conducted for human participants by Nakano et al. \cite{Nakano2010-sg}. The viewing time of the eyes by the G1 heads (median 42\%) was comparable to that observed in TD adults (median 43\%) but exceeded that of TD children (median 26\%). In contrast, attention to the mouth of the G1 heads (median 32\%) was two to three times higher than that in both TD adults (median 12\%) and children (median 19\%). Consequently, G1 heads devoted approximately 90\% of their viewing time to the face (median 92\%), which was significantly greater than those observed in TD adults (median 65\%) and children (median 66\%). These patterns are illustrated in a frame from clip No. 11 (Fig. \ref{fig:fig6}b). The G1-heads preferred to view faces of non-human animals as well (median 85\%, Fig. \ref{fig:fig6}c).

\begin{figure}[p]
\centering
\includegraphics[clip, width=\linewidth]{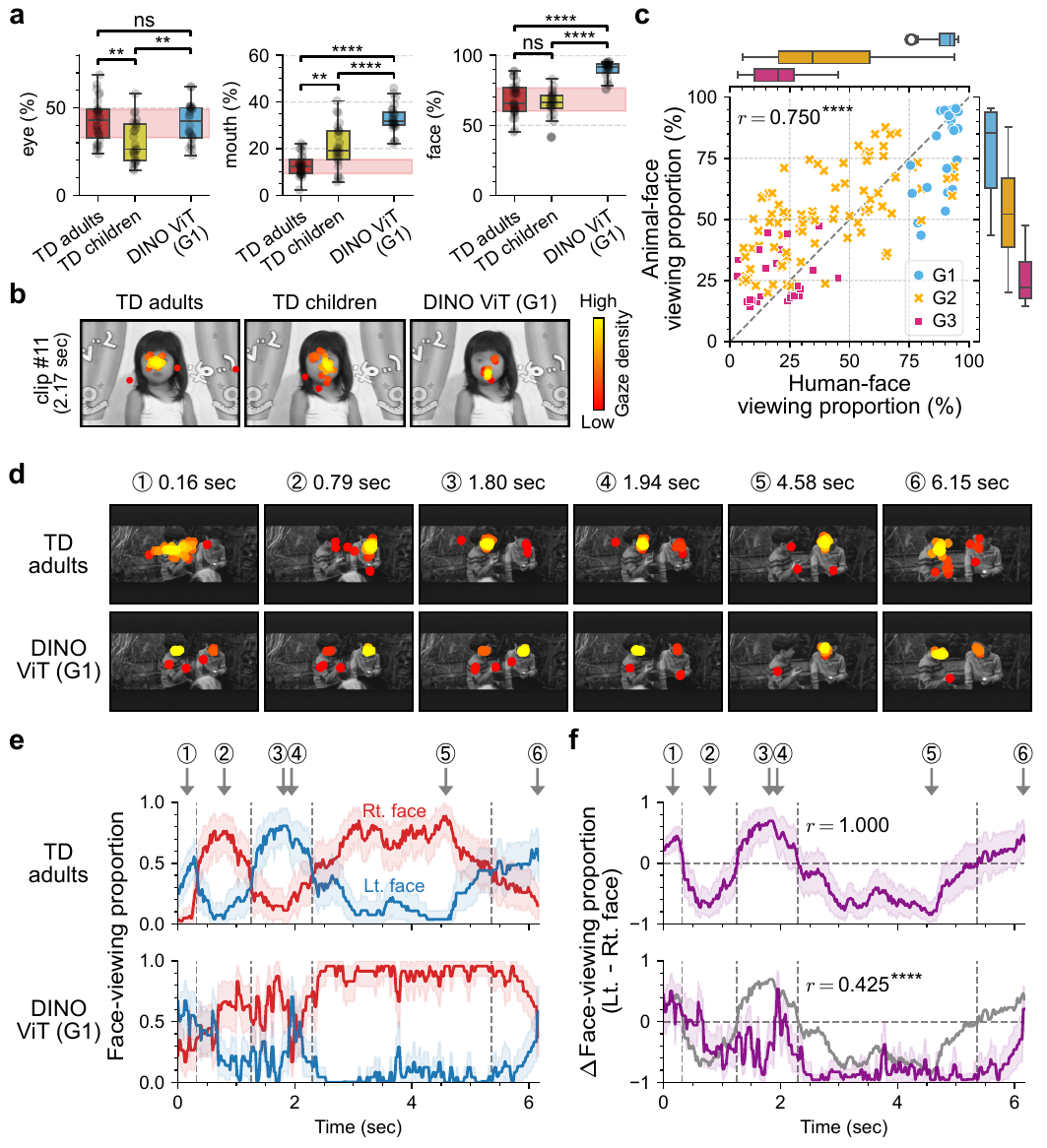}
\end{figure}

\begin{figure}[p]
\centering
\caption{\textbf{Comparison of face and face parts viewing proportions between human participants and G1 heads of DINO ViTs.} \textbf{a}, Group comparisons of viewing proportion for the eye (left), mouth (middle), and entire face (right). Clips 3, 7, and 11 of the N2010 dataset were used. Statistical significance is indicated as follows: **: $p<0.01$, ***: $p<0.001$, ****: $p<0.0001$, ns: $p\geq0.05$, using Wilcoxon rank-sum tests (two-sided, Bonferroni corrected). \textbf{b}, Examples of gaze positions focused on a face from a typical frame in clip 11 of the N2010 dataset. \textbf{c}, Scatter plot of viewing proportions for animal-faces versus human-faces across all G1, G2, and G3 heads. A strong Pearson correlation coefficient of 0.75 ($p < 10^{-26}$) is observed. Box plots display the distributions of viewing proportions for human faces (top) and animal faces (right). \textbf{d, e}, Comparison of the temporal dynamics of face-viewing proportions for two boys in clip No.4 between TD adults and DINO-ViTs. Note the synchronized attention switching between the two boys in TD adults, as shown by the complementary changes in red and blue curves in (\textbf{e}) for the right and left boys, respectively. Gaze positions at six representative frames, indicated by the arrows in (\textbf{e}), are displayed in (\textbf{d}). \textbf{f}, Differences between the two curves in (\textbf{e}), with the solid line indicating the mean and the shaded bands showing the 95\% confidence intervals. The Pearson correlation coefficient between the mean difference curves of TD adults and G1 heads is 0.425 ($p<10^{-14}$).}
\label{fig:fig6}
\end{figure}

\subsection{G1 heads partially replicated gaze switching between human faces}
In our previous studies \cite{Suda2015-he,Nakano2010-sg}, we observed that typically developing (TD) participants switched their gazes between characters in a remarkably synchronized manner. Figure \ref{fig:fig6}d presents the most thoroughly analyzed example from clip No. 4, where two boys are engaged in conversation. During this 6-second clip, the gazes of TD adults were initially centered around the boy on the left, who was closer to the center of the monitor (1). However, they quickly shifted their focus to the boy on the right at approximately 1 second (2). The gaze then switched back to the boy on the left around 2 seconds (3, 4), before returning to the boy on the right at 2.3 seconds for about 2.5 seconds (5), and finally settling back on the left boy at 5.6 seconds and remaining there (6). The graphs in Fig. \ref{fig:fig6}e display the face-viewing proportions for the two boys (red: right boy, blue: left boy), which alternated clearly, marking four distinct transitions (indicated by vertical dashed lines). By subtracting the viewing proportion of the right boy from that of the left boy, a single temporal profile was generated, crossing the time axis four times and delineating five distinct periods of dominant face viewing (Fig. \ref{fig:fig6}f).\par
The G1 heads of DINO ViTs failed to replicate these distinct transitions, instead consistently showing dominant attention to the boy on the right (Fig. \ref{fig:fig6}e and f, bottom row). According to Suda and Kitazawa \cite{Suda2015-he}, the switching process is well explained by a model that includes the size of each face, head motion, and mouth motion during vocalization. In this case, the face of the boy on the right, who was shown in a frontal view, appeared larger than the profile view of the boy on the left. Because motion signals from the face and mouth---critical cues for turn-taking---were not available in the current application of ViTs, the G1 heads were unable to reproduce the distinct gaze-switching patterns. However, there was a significant correlation between the single temporal profiles of TD adults and the G1 heads of DINO ViTs, with a coefficient of 0.425 (Fig. \ref{fig:fig6}f, $p < 10^{-14}$, comparison between means of each group). Since the size of the faces remained almost constant throughout the 6 seconds, it suggests that the G1 heads of DINO ViTs may have detected some subtle cue, other than size, that guides our attention to a face.

\subsection{G1 heads also responded to visual saliency}
\begin{figure}[p]
  \centering
  \includegraphics[clip]{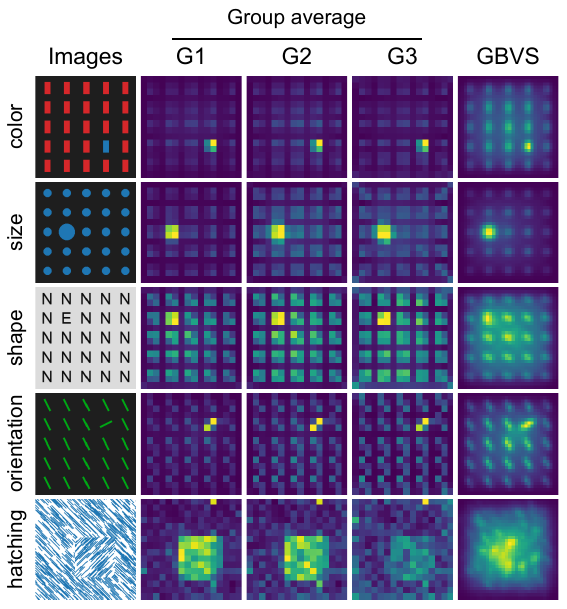}
  \caption{\textbf{Focus of attention on singletons within artificial stimuli.} Group-averaged attention maps for G1, G2 and G3 heads, alongside attention maps from GBVS (using DKL color, intensity and orientation channels) for five artificial psychological stimuli.}
  \label{fig:fig7}
\end{figure}

The results mentioned above demonstrate that the G1 heads exhibited a strong preference for faces. This raises the question of whether the G1 heads were merely face detectors. To address this, we also tested whether the attention of the G1 heads exhibited ``pop-out'' properties, meaning the ability to focus on an exceptional region (called singletons) based on physical properties (first-order features) such as orientation, color, size, and shape (Fig. \ref{fig:fig7}). The G1 heads were highly sensitive to visual saliency, characterized by color (first row), size (second row), shape (third row), and orientation (fourth row), performing as well as the GBVS model. Notably, the G1 heads were able to detect a pattern-defined object (a square distinguished from the background by the orientation of the hatching) involving a second-order feature, which the GBVS model failed to detect (fifth row). The attention of the G2 heads also displayed pop-out properties similar to those observed in the G1 heads. However, the G3 heads focused more on the background, though they also exhibited pop-out properties for first-order features.

%%%%%%%%%%%%%
%% Discussion %%
%%%%%%%%%%%%%
\section{Discussion}
We have revealed that self-attention from the class token of ViT exhibits a temporo-spatial pattern remarkably similar to that of human adults, especially when trained with DINO, but not when trained with conventional supervised learning with labels. Notably, a classical model based on feature-based attention, often referred to as a saliency model, yielded attention much less similar to human attention. These results indicate that not only the architecture of ViT but also the learning algorithm is essential for achieving human-like attention.\par
We also found that self-attention heads can be divided into three groups (G1, G2, and G3), each showing a preference for faces, bodies, and backgrounds, respectively. The G1 heads showed strong attention to human and animal faces but were also able to attend to singletons of artificial images with both first- and second-order features. This is remarkable because previous research, which used SL ViTs, has claimed that the attention mechanisms of ViTs do not exhibit attentional properties such as pop-out \cite{Mehrani2023-lg}. These results suggest that ViTs trained with DINO self-acquired attention comparable to the multifaceted attentional processes of humans.

\subsection{Is self-attention of ViTs biologically plausible?}
The biological plausibility of self-attention in ViTs, and in transformers in general, is often questioned. One concern is whether the brain’s visual areas have long-range connections comparable to the unrestricted range of self-attention in ViTs. The answer is yes---the brain has long-range connections within visual areas, allowing neurons to be influenced by both nearby and distant stimuli \cite{Angelucci2002-so,Sato2021-wq}\par
Another criticism is that transformers lack the top-down, or feedback, connections that exist in human brain networks, which play an essential role in generating top-down attention \cite{Lei2021-mu,Lindsay2020-vm}. Transformers may seem to lack recurrent connections, but ViTs are equipped with the class token, which aggregates information from all the input tokens. The class token sends queries to these tokens, and those queries are correlated with the keys in the input tokens to form attention weights. The weighted sum of values is then sent back to the class token. The biological plausibility of dot products between queries, keys, and values has also been questioned, but several researchers have proposed ways to achieve these by using biologically plausible components \cite{Ellwood2024-fz,Kozachkov2023-wo,Krotov2021-zm}. Thus, the entire process is, in principle, comparable to widespread backward projections in actual neural networks and forward projections that integrate information from lower layers to higher layers. From this, we argue that top-down connections do exist in ViT, from the class token to all the others, capturing the essence of top-down attention. \par
It could still be argued that the class token does not send signals back across multiple layers to the input layer. However, the class token’s attention in a single layer still resembles human attention. This raises the question of how far back human top-down attention extends in the brain. Given the frequent and rapid shifts in retinal images due to saccades, it seems unlikely that top-down signals must return to the primary visual cortex to finalize attention. Our finding that class token attention in layers 9 and 10 of DINO ViT-12 closely mirrors human attention (Fig. \ref{fig:fig4}a) suggests that a more localized top-down process within higher cortical layers may be sufficient to explain top-down attention.

\subsection{Why do DINO ViTs exhibit human-like attention?}
The question arises: why do DINO ViTs exhibit attention patterns much more similar to human attention than SL ViTs? We speculate that two key aspects of DINO's self-supervised learning method reflect a learning process akin to that of a newborn.\par
DINO's loss function, $\mathcal{L}_\mathrm{DINO}$, is the sum of the Kullback-Leibler (KL) divergence between the probability distribution produced by the teacher ViT, $P_\mathrm{t}$, and that by the student ViT, $P_\mathrm{s}$, and the entropy of $P_\mathrm{t}$:
\begin{equation}
\mathcal{L}_\mathrm{DINO}\coloneqq P_\mathrm{t}^\top \log(P_\mathrm{t}/P_\mathrm{s}) - P_\mathrm{t}^\top \log P_\mathrm{t}
\end{equation}
First, DINO learning minimizes the KL divergence between the student and the teacher. The student ViT observes many different portions of an original image, similar to how the human eye makes saccades, while the teacher ViT observes a larger portion of the image. This process leads to view-invariant representation learning, promoting visual stability across saccades.\par
Second, the DINO minimizes the entropy of the teacher ViT’s output, $P_\mathrm{t}$, while its ``centering'' procedure flattens the mean of $P_\mathrm{t}$, effectively maximizing the entropy of its prior distribution. Together, these procedures allow DINO to maximize Shannon information the difference between the entropy of the prior and posterior distributions, before and after observing a new image. \par
We might call this the ``DINO information maximization principle,'' though it should not be confused with conventional mutual information maximization like InfoMax \cite{Bell1997-rq,Linsker1988-uy} or DeepInfoMax \cite{Hjelm2019-rl}. Traditional InfoMax maximizes mutual information between input data and its neural network representation, requiring both input and output distributions. DINO, on the other hand, uniquely focuses only on the output distribution.\par
It could be argued that the conventional SL method also maximizes information about predefined labels from an image. However, a fundamental difference exists between DINO and SL: SL deals with a fixed set of 1,000 labels, leaving no flexibility for the ViT to improve or adapt those labels, even if they lack optimal informativeness. DINO, by contrast, operates with an output dimension of up to 65,536, allowing the ViT to autonomously organize this vast space to optimally represent the complexity of the world. \par
Now, consider which process is more akin to a newborn’s learning experience. A newborn knows little about how the visual world is organized, let alone the classifications defined by human experts. Maximizing information through observation, as DINO does, seems more ecological for a baby adapting to an unpredictable environment. Indeed, the mammalian visual system has sensitive periods during which experiences shape sensory filters in primary sensory cortices \cite{Berardi2000-wz,Hubel1963-si,Knudsen2004-hr}.\par 
Although DINO uses non-physiological components---like the student-teacher system---it is tempting to speculate that the brain may have evolved a more efficient, autonomous mechanism for maximizing information without relying on such structures. The remarkable similarity between DINO ViTs’ attention and human attention suggests that DINO’s information maximization could reflect a fundamental principle of learning in a novice brain.

\subsection{How do G1 heads learn to focus on a face?}
A striking feature of G1 heads is their keen interest in focusing on faces within a scene. Their viewing proportion of faces was as high as 90\%, surpassing the 60-70\% observed in TD adults and children (Fig. \ref{fig:fig6}). Suda and Kitazawa \cite{Suda2015-he} demonstrated that a gaze scanpath, generated by randomly selecting one of multiple faces in a scene, can significantly differ from the actual scanpaths generated by TD adults. The fact that the scanpaths of G1 heads were comparable to those of TD adults suggests that G1 heads chose the same primary face in the scene as TD adults. How could the G1 heads identify an appropriate face without being taught what a face is?\par
To address this issue, we examined how many images in the ImageNet-1k dataset, used for training DINO ViTs, contained a human face or a human. Of 1,000 images randomly drawn from the dataset, 159 contained some parts of humans or human-like figures (e.g., images shown with red rectangles in Supplementary Fig. \ref{fig:fig_sup3}). There was a wide variety in the size and numbers of humans in these pictures, but more than half (98/159) contained at least a recognizable face. In addition, the dataset featured a wide variety of animals (388 out of 1,000) with recognizable heads and faces. Taken together, nearly half of the dataset images contained a face, whether human or non-human animal. Consider obtaining information by discriminating between the presence and absence of elements in an image. The maximal information of one bit is obtained when an object appears in an image with a probability of 1/2. We speculate that DINO ViTs autonomously developed a face detector because faces appeared with an approximate probability of 1/2 in their virtual world of ImageNet. This speculation is supported by the fact that the G1 heads paid attention to both human and non-human animal faces (Figs. \ref{fig:fig5} and \ref{fig:fig6}). 

\subsection{Implications of G1, G2, and G3 attention heads in human visual perception}
It was unexpected that self-attention heads in the critical layers of DINO ViTs clustered into three groups rather than two, given that conventional literature in visual psychology often emphasizes dichotomies, such as figure-ground segregation \cite{Wagemans2012-qx}. In this context, G2 aligns with figures and G3 with the ground. G2 attention is evenly distributed over human and animal bodies or main objects, maintaining precise boundaries (e.g., Fig. \ref{fig:fig5}). This boundary precision suggests that DINO ViTs may address the border ownership problem, typically managed by lower visual areas like V2 \cite{Qiu2005-tg}. Additionally, G2, comprising 60\% of the critical attention heads (Fig. \ref{fig:fig4}d), should be useful for ``semantic segmentation'' \cite{Caron2021-qj,Shelhamer2014-lm}, which assigns a label to each segmented area. \par
The role and neural basis of G3 also merit consideration. A recent study has indicated that many neurons in the precuneus---a major hub of the neural network---encode the retinotopic location of a large background \cite{Uchimura2024-qu}, closely paralleling the role of G3 heads. Furthermore, significant numbers of the precuneus neurons encode both retinotopic information of a dot (figure) and allocentric information of the figure relative to the background. Taken together, these precuneus neurons seem to include both G2- and G3-like functions, representing stabilized allocentric information that remains view-invariant across saccades. \par
If G2 and G3 attention heads explain most findings in figure-ground segregation, what role do G1 heads play? In typical psychological experiments with only one main figure, G2 and G3 heads may suffice. However, natural scenes often contain multiple figures, and the visual attention system continuously selects one as the focal point. G1 heads likely serve this purpose by identifying the ``center'' within figures represented by G2 heads. Neural correlates of G2 heads, if they exist, may act as a ``reservoir'' of figures from which G1 heads select the central focus.\par
Several research studies has shown general correspondence between layers of convolutional artificial neural networks and the hierarchies of visual cortical areas \cite{Margalit2024-bn,Yamins2014-fn,Yamins2016-vz}. Consequently, neural correlates of G1, G2, and G3 heads, found in layers 9 and 10 of DINO ViT-12, are likely to be found at high levels in the visual hierarchy from the retina to the hippocampus as illustrated by Felleman and Van Essen \cite{Felleman1991-fv}. Future research will examine these neural correlates in the human brain by comparing the effects of selective ablation or inhibition in both DINO ViTs and biological networks.

%%%%%%%%%%%%%
%% Methods %%
%%%%%%%%%%%%%
\section{Methods} 
\subsection{Video stimuli and eye tracking data}
The video stimuli and human eye tracking data used in this study were derived from Nakano, et al. \cite{Nakano2010-sg}, N2010, and Costela and Woods \cite{Costela2019-ba}, CW2019. The N2010 data were obtained from 104 participants (27 adults and 25 children with typical development, and 27 adults and 25 children with Autism Spectral Disorder, ASD) while they were viewing a 77-second-long video clip, consisting of 12 short video clips taken from TV programs and movies, each lasting for 5-6 s and featuring one or more human characters. Gaze positions were measured with the Tobii X50 system (Tobii Technology AB), with a sampling frequency of 50 Hz. \par
We used a part of the CW2019 data (\url{https://osf.io/g64tk/}), obtained from 63 adult participants while they were viewing ``Hollywood'' video clips, each lasting for approximately 30 seconds, using the EyeLink 1000 system (SR Research Ltd.), with a sampling frequency of 1,000 Hz. The dataset contained 200 videos across three genres: ``drama/other'' featuring human actors ($n = 120$), ``cartoon/animation'' featuring human-like characters ($n = 40$), and ``documentary/nature'' films featuring animals or natural scenes ($n = 40$). It should be mentioned that each participant viewed approximately 40 of the 200 video clips.

\subsection{Vision Transformers (ViTs)}
\paragraph{Architecture.}
The architecture of ViTs used in this study was based on the model called ``DeiT-S''\cite{Touvron2021-yb} or ``ViT-S/16''\cite{Caron2021-qj} which has the following parameters: patch size $n_p=16$ pixels, embedding dimensions  $D_\textrm{emb} =384$, number of heads $n_h=6$, and 12 transformer layers $L=12$. Additionally, we developed two smaller ViT models with $L=4$ or $8$ for comparison.\par
A ViT processes an image first partitioning it into $N$ patches, each of size $n_p^2$. Each patch is then linearly transformed into a $D_\textrm{emb}$-dimensional embedding vector (in this case, 384-dimensional), referred to as a patch token. This transformation is performed using 384 learnable linear filters. After training, many of these filters resemble Gabor-like filters, reminiscent of the receptive fields of V1 neurons \cite{Dosovitskiy2021-tp}. Position information is then added to each patch token. Notably, ViTs include a class token (\texttt{[CLS]}), which is appended to the beginning of the $N$ patch tokens (Fig. \ref{fig:fig1}). This token aggregates information across the entire set of $N$ patches. \par
The resulting input is an $(N+1)\times D_\textrm{emb}$-dimensional matrix $\mathbf{X}$, where each row is fed into each of $N+1$ modules of the first layer of the transformer encoder (Fig. \ref{fig:fig1}). Each of $N+1$ modules of the transformer layer has $n_h$ ``attention heads'', each of which combines information across the $N+1$ tokens in parallel: the matrix $\mathbf{X}$ is divided into $n_h$ sub-matrices $\mathbf{X}_i\in \mathbb{R}^{(N+1)\times D_h}$, where $\mathbf{X}=\left[\mathbf{X}_1,\mathbf{X}_2,\ldots,\mathbf{X}_{n_h}\right]$ and $D_h=D_\textrm{emb}/n_h$.  
Each attention head $i\ (i=1,\ldots,n_h)$ mixes information across the $N+1$ tokens in $\mathbf{X}_i$ using self-attention mechanisms as follows:
\begin{equation}
    \textrm{Attention}(\mathbf{Q}_i,\mathbf{K}_i,\mathbf{V}_i)\coloneqq\textrm{Softmax}\left(\frac{\mathbf{Q}_i\mathbf{K}_i^\top}{\sqrt{D_h}}\right)\mathbf{V}_i\label{eq:attn}
\end{equation}
where $\mathbf{Q}_i=\mathbf{X}_i\mathbf{W}_i^\textrm{Q}, \mathbf{K}_i=\mathbf{X}_i\mathbf{W}_i^\textrm{K}, \mathbf{V}_i=\mathbf{X}_i\mathbf{W}_i^\textrm{V}\ \left(\mathbf{Q}_i,\mathbf{K}_i,\mathbf{V}_i\in \mathbb{R}^{(N+1)\times D_h}\right)$ are the query, key and value matrices, respectively, and $\mathbf{W}_i^\textrm{Q}, \mathbf{W}_i^\textrm{K}, \mathbf{W}_i^\textrm{V} \in \mathbb{R}^{D_h \times D_h}$ are the learnable linear transformation matrices. Equation (\ref{eq:attn}) shows that the $j$-th row of $\textrm{Softmax}\left(\mathbf{Q}_i\mathbf{K}_i^\top/\sqrt{D_h}\right) \in \mathbb{R}^{(N+1)\times(N+1)}$, referred to as the attention weight matrix, defines a set of attention weights for the head $i$ in the $j$-th token. These weights determine how each column of the value matrix (of dimension $N+1$) is weighted and averaged. Thus, the first row of the attention weight matrix gives the attention weights for the class token. \par
The self-attention map of the class token in the $i$-th head, $\mathbf{a}_i^{\texttt{[CLS]}}$, is given by:
\begin{equation}
\mathbf{a}_i^{\texttt{[CLS]}}\coloneqq\left[\textrm{Softmax}\left(\mathbf{q}_i^{\texttt{[CLS]}}\mathbf{K}_i^\top/\sqrt{D_h}\right)\right]_{2:N+1}\in\mathbb{R}^N
\end{equation}
where $\mathbf{q}_i^{\texttt{[CLS]}}\in\mathbb{R}^{1\times D_h}$ is the first row of $\mathbf{Q}_i$, and $[\cdot]_{2:N+1}$ means taking the second to $(N+1)$-th elements of the vector. This produces an attention map over the $N$ patch tokens, which is then reshaped into two dimensions with an aspect ratio that matches the input image. Henceforth, when we refer to an attention map, we specifically mean the attention map of the class token over the $N$ patch tokens. 
In the first layer, the outputs from the six attention heads are concatenated and passed through additional processing using a multilayer perceptron (MLP) and skip connections, resulting in an $(N+1)\times D_\textrm{emb}$-dimensional matrix. The output matrix of the first layer is fed into the second layer. This process is repeated across all layers up to the final $L$-th layer. The final output is a $D_\textrm{emb}$-dimensional vector derived from the class token, which is then passed through an MLP, known as the projection head, to generate the final output. For supervised learning, the output dimensionality was $D_\textrm{out}=1,000$, whereas for DINO-based learning it was $D_\textrm{out}=65,536$.

\paragraph{Acquisition of gaze positions from ViTs.} 
We resized video frames or images so that their dimensions (length and width) were multiples of $n_p$ for input into ViTs. The resized video frames were presented independently to the trained ViTs. Attention maps obtained from the trained ViTs were up-sampled to match the size of the input image using nearest-neighbor interpolation and then smoothed using a box blur with a kernel size of $2n_p$. Following the winner-take-all principle, we deterministically defined the peak positions of the attention map as the gaze positions of each attentional head (Fig. \ref{fig:fig1}). For comparison purposes, these gaze positions were then mapped to the coordinates of the original image.

\paragraph{Training.}
We trained a total of 36 ViTs, with six models for each structure ($L = 4,8$, and 12), using both supervised learning and self-supervised learning (DINO). For both training approaches, we used the training set from the ImageNet-1k (ILSVRC2012) dataset \cite{Russakovsky2015-we}, consisting of 1,281,167 images labelled across 1,000 classes. \par
In the supervised learning setup, we trained the ViTs to minimize the cross-entropy loss between the output of the projection heads and the 1000 labels. We used the training code from Touvron, et al. \cite{Touvron2021-yb}, available at: \url{https://github.com/facebookresearch/deit}. The models were trained using default parameters, without distillation. \par
For the DINO training, we used the code from Caron, et al. \cite{Caron2021-qj}, available at: \url{https://github.com/facebookresearch/dino}. We trained the model for 300 epochs (the same as in supervised learning) using default hyperparameters.\par
DINO training, introduced by Caron, et al. \cite{Caron2021-qj}, is unique in that it does not rely on pre-defined labels. Instead, the ViT learns to discover ``optimal classifications'' of visual images, leveraging a large output dimension of 65,536. In DINO, two ViTs with identical architectures are employed: one acting as the student and the other as the teacher. Both models output a probability distribution over the 65,536 bins, denoted $P_\textrm{s}$ (student), and $P_\textrm{t}$ (teacher). The loss function, $\mathcal{L}_\textrm{DINO}$, is the cross-entropy between $P_\textrm{s}$ and $P_\textrm{t}$, which can be decomposed into the entropy of $P_\textrm{t}$ and the Kullback-Leibler (KL) divergence between $P_\textrm{t}$ and $P_\textrm{s}$:
\begin{equation}
    \mathcal{L}_\textrm{DINO}\coloneqq - P_\mathrm{t}^\top \log P_\mathrm{s}=P_\mathrm{t}^\top \log(P_\mathrm{t}/P_\mathrm{s}) - P_\mathrm{t}^\top \log P_\mathrm{t}
\end{equation}
Three key mechanisms are crucial in DINO learning. First, the input images are augmented by cropping smaller regions (local views) and larger regions (global views). All the crops are shown to the student, while only the global views are shown to the teacher. Minimizing the KL divergence between $P_\textrm{t}$ and $P_\textrm{s}$ encourages the model to learn view-invariant representations. \par
Second, a centering procedure is applied to the teacher’s output to prevent model collapse, where the teacher would otherwise produce the same output for all inputs. This centering ensures that the model uses the full 65,556 dimensions for classification. By assuming a uniform prior distribution across the bins, minimizing the entropy term maximizes the Shannon information, defined as the difference between the entropy before and observing a new image. \par
Third, the update rules of the parameters differ between the student and teacher models. The student’s parameters are updated directly by minimizing the loss $\mathcal{L}_\textrm{DINO}$, while the teacher’s parameters are updated with the exponential moving average (EMA) of the student’s parameters. This EMA approach introduces a model ensemble effect, enhancing stability and gradually leading to convergence of both models' parameters. \par
In summary, DINO training encourages the ViT to discover optimal classifications, maximizing the information gained from observing new images. 

\paragraph{Utilization of ``official models''.}
In addition to the 36 ViTs we trained ourselves, we used publicly available two pre-trained 12-layer models, one trained with supervised learning (DeiT-S) and the other with DINO (ViT-S/16). We refer to these as ``official models''. The trained weights of the official models are available from the links mentioned above.

\paragraph{Environment of model training.}
The ViT models were trained on a workstation equipped with an AMD EPYC 7513 CPU, 512 GB of RAM, and eight NVIDIA A40 GPUs. The software used for training was Python 3.8.8 with PyTorch 1.8.1.

\subsection{Graph-Based Visual Saliency model}
A Graph-Based Visual Saliency (GBVS) model \cite{Harel2006-wp} was used as a control. This model includes six feature channels—color, intensity, orientation, contrast, flicker and motion—to capture various aspects of saliency. For the color channel, we used DKL (Derrington-Krauskopf-Lennie) color space, which aligns with human visual perception. Gaze positions in the GBVS were defined by the peak saliency values in the attention maps, similar to the method used for ViTs.

\subsection{MDS analysis for gaze positions}
To quantify differences and similarities in the temporal pattern of gaze movements among human participants, ViTs, and GBVS models, we computed pairwise distances between gaze positions at each time point across the video clips \cite{Nakano2010-sg}. Prior to calculation, gaze position sampling rates were adjusted to 50 Hz for the N2010 data and 720 Hz for the CW2019 data using nearest-neighbor interpolation.\par

We defined the distance $d_{ij}$ between two time series of gaze positions, $\mathbf{v}_i(t), \mathbf{v}_j(t) \in \mathbb{R}^{2}\ (t=0, \ldots, T,\ i\neq j)$ from the $i$-th and the $j$-th participant (or model) as follows: 

\begin{equation}
    d_{ij} \coloneqq \mathop{\texttt{nanmedian}}\limits_{t\in [0, T]}\left(\|\mathbf{v}_i(t) - \mathbf{v}_j(t)\|_2\right)
\end{equation}
where, $\|\cdot\|_2$ denotes the Euclidean norm of a vector, and $\texttt{nanmedian}(\cdot)$ is a function that calculates the median while ignoring missing values (NaNs). 

For the CW2019 data, distances $d_{ij}$ were computed separately for each video clip and normalized by the median distance, as each participant viewed a different set of clips. The resulting distance matrices were then averaged across clips. Missing values in $d_{ij}$ were interpolated using the following formula:
\begin{equation}
 \hat{d}_{ij} = \hat{d}_{ji} \coloneqq (\texttt{nanmedian}_k(d_{ik}) + \texttt{nanmedian}_k(d_{kj}))/2   
\end{equation}
First, we applied metric multidimensional scaling (MDS) on the distance matrix between human participants to establish a standard MDS space where the origin represents the general gaze profile of human participants in general. Data from ViTs and GBVS models were then plotted relative to these landmarks using the following formula: 
\begin{equation}
    \mathbf{u}_k = \mathop{\textrm{arg min}}\limits_{\tilde{\mathbf{u}}_k} \left[\sum_{i\in \mathcal{S}}\left(\|\mathbf{u}_i - \tilde{\mathbf{u}}_k\|_2 - d_{ik}\right)^2\right] \label{eq:mds_model}
\end{equation}
where $\mathcal{S}$ is the set of landmark data (human participants), $\mathbf{u}_i$ is the embedding vector of the $i$-th participant in the MDS space, and $\mathbf{u}_k$ is the embedding vector of the $k$-th model. We performed MDS analysis and nonlinear optimization of equation (\ref{eq:mds_model}) using MATLAB R2023a (MathWorks, Natick, MA, USA), specifically with the functions ``mdscale'' and ``lsqnonlin'' with ``MultiStart''.\par
To quantify deviation from standard human gaze behavior, we defined the MDS distance, representing the distance from the median of the human data. While we used two-dimensional MDS for visually representing similarities and differences in gaze data (e.g., Fig. \ref{fig:fig2}a), we increased the dimensionality to 32 for formal MDS distance calculations to minimize underestimation of distance. Our analysis indicated that using two dimensions could lead to an underestimation of up to 9\%, whereas a 32-dimensional space reduced this error to less than 1\%. 

\subsection{Hierarchical clustering of self-attention heads}
We applied hierarchical clustering to classify self-attention heads based on the attention maps they generated. The distance matrix was calculated by averaging the cosine distances (one minus cosine similarities) of the attention maps across the N2010 video frames. This matrix was then used for hierarchical clustering. To determine the optimal number of clusters, we used 24 clustering indices provided by the NbClust package (ver. 3.0.1)\cite{Charrad2014-kw} in R 4.3.2. To calculate these cluster indices, we also used a matrix of attention maps for N2010 video frames concatenated by each self-attention head and compressed into 128 dimensions with principal component analysis (PCA).

\subsection{Viewing proportion analysis}
We calculated the viewing proportion, a metric raging between 0 and 1, indicating whether a human or an artificial model is looking at a specific target, such as the eyes, mouth, or face \cite{Nakano2010-sg}. The viewing proportion for the $i$-th target at time point $t$, denoted by $\omega_i (t)$, is given by:
\begin{align}
    \tilde{\omega}_i(t)&\coloneqq\exp\left({\frac{-\|\mathbf{v}(t)-\boldsymbol{\xi}_i(t)\|^2}{2\sigma^2}}\right)\\
    \omega_i(t)&\coloneqq\frac{\tilde{\omega}_i(t)}{\max(1,\ \sum_j \tilde{\omega}_j(t))}
\end{align}
where $\mathbf{v}(t)\in\mathbb{R}^2$ is a gaze position, $\boldsymbol{\xi}_i(t)\in\mathbb{R}^2$ is the position of the $i$-th target, and $\sigma$ is the standard deviation parameter of the Gaussian function, set to 30 pixels. When the $i$-th target is absent, $\omega_i (t)$ is set to zero.
The average viewing proportion of $\omega_i (t)$ over time, calculated by summing $\omega_i (t)$ and then dividing by the total time the $i$-th key point was present, indicates the proportion of time spent viewing the $i$-th target. We did not compute the viewing proportions of self-attention heads in ViTs directly from the attention map, to ensure consistency in analytical methods between data from human participants and models.\par
For human face-viewing proportions, we used clips No. 3, 7, and 11 from the N2010 data (a total of 571 frames), each manually annotated with target coordinates (eyes, mouth, nose, ears, and hands) for every frame. The face-viewing proportion was defined as the sum of viewing proportions for the eyes, mouth, nose, and ears. 
For non-human animal face-viewing proportions, we used the Animal Parts Dataset \cite{Novotny2016-yj} containing eye and foot annotations across 14,711 vertebrate images from the Imagenet-1k dataset. We selected 875 images based on the following criteria: images were from the Imagenet-1k validation set (not used in model training), had visible faces, contained no humans or multiple faces, and depicted biological (non-artificial) objects. We manually annotated mouth positions for these images; for animals with elongated beaks or mouths (e.g., birds, crocodiles) the mouth position was set approximately at the center. For animal faces, the face-viewing proportion was defined as the sum of viewing proportions for the eyes and the mouth. To obtain the gaze positions of the ViTs, the dataset images were cropped and resized to a resolution of 256 $\times$ 256 pixels.

\section*{Data availability}
The video stimuli and eye-tracking data from N2010 \cite{Nakano2010-sg} are available upon request. The data from CW2019 \cite{Costela2019-ba} are openly available (\url{https://osf.io/g64tk/}).

\section*{Code availability}
The code is publicly available at: \url{https://github.com/KitazawaLab/vit-human-attention-comparison}.

\section*{Acknowledgments}
This work was supported by KAKENHI 23K17462, 21H04896, and 18H05522 from the Japan Society for the Promotion of Science (JSPS) to SK. Portions of Figure 1 were created using BioRender (\url{https://www.biorender.com/}). Language checking was conducted using ChatGPT-4.

\section*{Author contributions}
T.Y., H.A. and S.K. designed the study. T.Y. and H.A. conducted model training and data analysis. T.Y. and S.K. wrote the manuscript. All authors reviewed the manuscript.

\section*{Competing interests}
The authors declare no competing interests.

% citation
\bibliographystyle{naturemag}
\bibliography{references}  %%% Remove comment to use the external .bib file (using bibtex).

\clearpage
\section*{Supplementary material}
\label{sec:sample:appendix}
\subsection*{Supplementary figures}
\renewcommand{\theHfigure}{A\arabic{figure}}
\renewcommand{\figurename}{Supplementary Figure}
\setcounter{figure}{0}
\begin{figure}[H]
  \centering
  \includegraphics[clip, width=\linewidth]{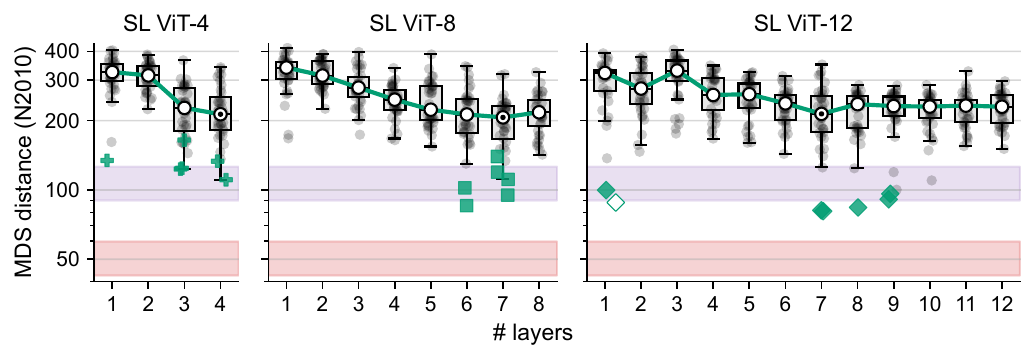}
  \caption{\textbf{Layer-wise MDS distance distribution in SL ViTs (N2010 dataset). } MDS distance of all self-attention heads at each layer in the SL ViTs. As in Fig. \ref{fig:fig4}a, the white circle represents the median MDS distance per layer, and the $\odot$ markers indicate the layer with the lowest median value. The best head in each model is shown by a green (SL) marker. Bands are consistent with those in Fig. \ref{fig:fig2}c.}
  \label{fig:fig_sup1}
\end{figure}

\begin{figure}[p]
  \centering
  \includegraphics[clip, width=\linewidth]{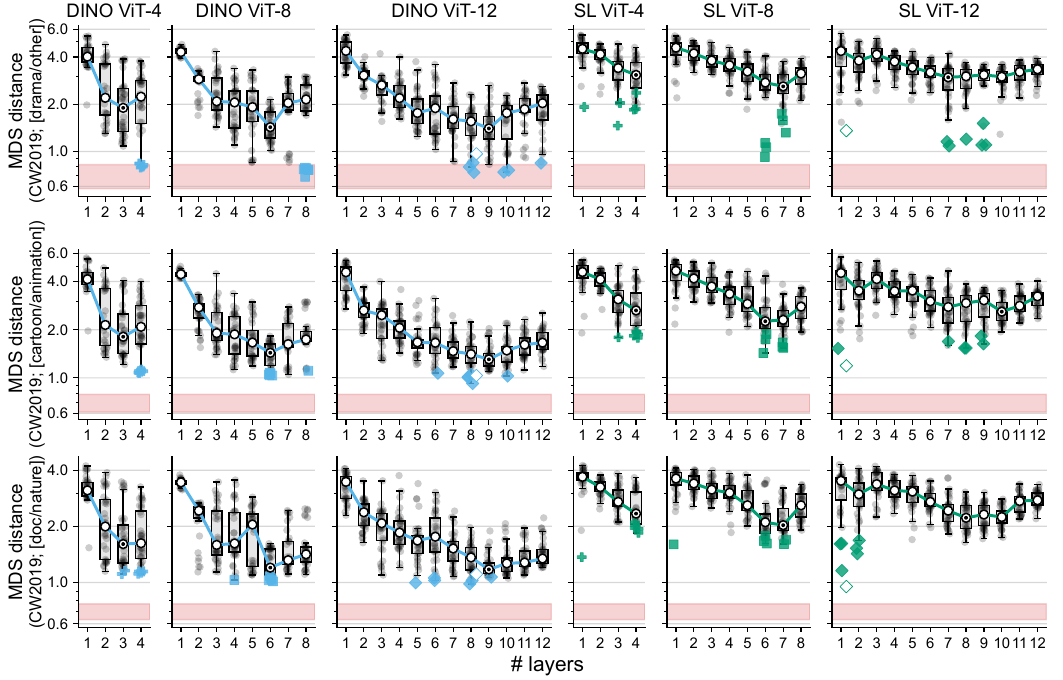}
  \caption{\textbf{Layer-wise MDS distance distribution in DINO and SL ViTs (CW2019 dataset).} The left and right panels show layer-wise MDS distances for DINO and SL ViTs, respectively, using CW2019 data. From top to bottom, results are shown for the ``drama/other'', ``cartoon/animation'', and ``documentary/nature'' video genres. As in Fig. \ref{fig:fig4}a, the white circle represents the median MDS distance per layer, and the $\odot$ markers indicate the layer with the minimum median value. The first best head in each model is indicated by a cyan (DINO) or green (SL) symbol. Bands are consistent with those in Fig. \ref{fig:fig3}c.}
  \label{fig:fig_sup2}
\end{figure}

\begin{figure}[p]
  \centering
  \includegraphics[clip, width=\linewidth]{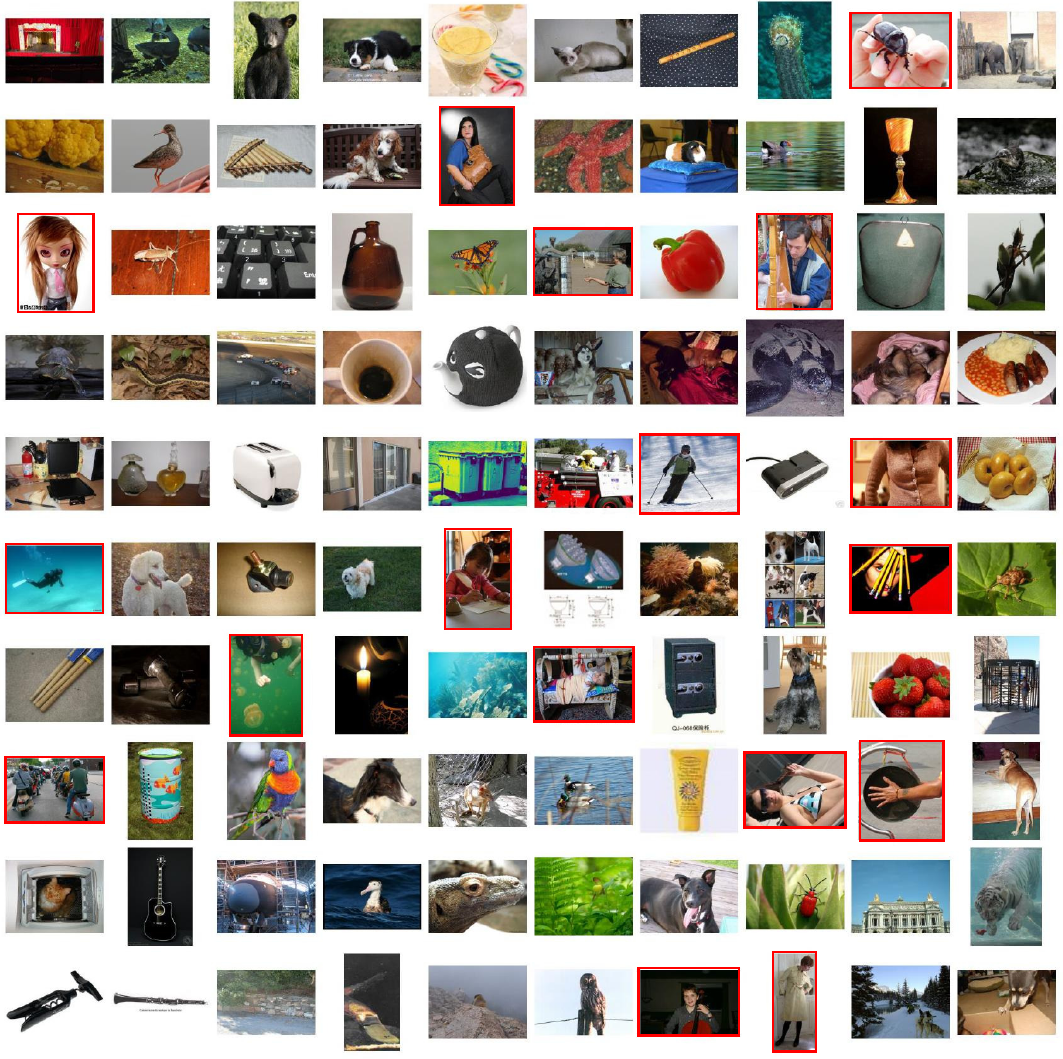}
  \caption{\textbf{Examples of images of humans or human-like figures in ImageNet-1k.} A set of 100 images randomly selected from the ImageNet-1k dataset, with red rectangles highlighting human or human-like figures.}
  \label{fig:fig_sup3}
\end{figure}
\end{document}